# Identifying a Level-up Pathway for AI-assisted Counterspeech through Elaboration

Han Li[1*], Inhwan Bae[1], Natalie Bazarova[1], Drew Margolin[1]

[1]Department of Communication, Cornell University, Ithaca, 14853, NY, USA.

*Corresponding author. E-mail: hl2564@cornell.edu

Contributing authors: ib295@cornell.edu; nnb8@cornell.edu; dm658@cornell.edu

## Abstract

Given the profound societal impact of vaccine-skeptical content on social media, community-driven counterspeech has emerged as a promising participatory response to contest and curb such objectionable content. Yet crafting effective counterspeech remains challenging for ordinary users, limiting their willingness and ability to engage constructively. We designed and evaluated three generative AI-assisted counterspeech writing systems that vary by assistance stage (co-writing vs. re-writing) and mode (guided vs. unguided) to support lay users' responses to vaccine-skeptical content. We ask whether AI can help users craft counterspeech perceived as both effective and authentic, which forms of AI support work best, and through what mechanisms. In a randomized controlled trial with social media users, participants wrote counterspeech responses to both statistical and narrative vaccine-skeptical content. Across evidence types, AI-assisted writing increased perceived counterspeech effectiveness while largely preserving authentic self-expression, and perceived effectiveness was the strongest predictor of willingness to counterspeak publicly. AI's primary benefit was facilitating more elaborate writing, producing messages that were more informative, analytical, and lexically sophisticated. These findings suggest a level-up pathway for AI-assisted, community-driven counterspeech, which helps cultivate more effective and motivated counterspeakers, contributing to higher-quality public discourse on pressing societal issues.

## 1 Introduction

In the aftermath of COVID-19, the proliferation of vaccine misinformation online has become a pressing global public health crisis, fueling vaccine hesitancy[1,2], eroding public trust in government, science, and public institutions[3,4] and drawing questions of collective health into the deeper currents of political and societal polarization[5,6]. To date, scholarly and practical efforts to address vaccine misinformation have largely focused on content that can be clearly and explicitly identified as false or misleading[1], with platform-driven moderation, sometimes in partnership with third-party fact-checking organizations, playing a central role in removing and labeling misinformation or disinformation to curb its circulation and influence[7-9].

Although these top-down approaches have produced measurable impacts, they are bounded by the very logic that makes them practically feasible. They work best when problematic content can be clearly detected, verified, and adjudicated as misinformation or disinformation, allowing the platform or moderator to take a clear, declarative stance. Yet much of the vaccine-related discourse that circulates most visibly on social media does not take the form of demonstrably false claims. In a sweeping study, researchers found that vaccine-skeptical content – content that raises questions about vaccines without being factually inaccurate – had 46 times greater impact in fostering vaccine skepticism than flagged misinformation[1]. Strikingly, the same study found that unflagged, factually accurate yet skeptical content can vastly outpace falsehoods in reach. These findings demonstrate ambiguous and less easily verified discourse that casts vaccines as harmful to health is a critical, but understudied, challenge.

For this type of vaccine-skeptical content, top-down interventions are not only ill-suited, they also risk backfiring when perceived as censorship, suppression, or institutional overreach[10-12]. In response, an alternative approach has gained increasing attention: community-driven counterspeech. Rather than relying on platforms or institutional authorities to identify and correct problematic content, counterspeech centers the voluntary participation of ordinary users who speak up against objectionable content online[13,14]. X's (former Twitter) Community Notes, represents one such endeavor. By mobilizing crowdsourced efforts to collaboratively add contexts to potentially misleading posts, Community Notes illustrate how distributed, algorithmically-managed, bottom-up correction may complement institutional forms of moderation. Prior research suggests that exposure to Community Notes can reduce the subsequent spread of misleading posts by 61% and increase the odds that users delete their misleading posts by 94%[15].

However, the promise of community-driven counterspeech rests on a crucial premise: the presence of motivated and well-informed counterspeakers who are both willing to enter these contested spaces and capable of responding in ways that are credible and effective, qualities which research indicates are rare[16,17]. Critically, these barriers compound one another: "how-to" challenges may further undermine upstanders' perceived self-efficacy and responsibility, both key factors underlying the motivation to counterspeak[18].

Moreover, effective counterspeech requires more of users than simply calling out misleading content. Pointing out that a claim is problematic is unlikely to produce durable effects unless the response also provides stronger arguments, clarifying contexts, or alternative evidence that can fill the interpretable gaps through which doubt takes hold[19,20]. This means that counterspeakers must perceive and respond to the evidentiary logic through which misleading content appears credible or compelling.

We ask whether artificial intelligence (AI) can help potential counterspeakers overcome these barriers to speaking up against vaccine-skeptical content. We developed and tested three AI-assisted writing systems designed to help ordinary users craft effective counterspeech in response to real-world vaccine-skeptical posts. By varying both the writing stage at which AI assistance is provided (co-writing vs. rewriting) and the mode of assistance (unguided vs. guided), we identify what lay counterspeakers seek from AI support, what the AI assistance actually contributes to their messages, and whether the assistance results in counterspeech that is perceived as more effective while preserving user's authentic sense of self-expression.

Vaccine-skeptical content often draws on two distinct evidentiary forms. The first is statistical evidence, in which numbers, statistics, or scientific findings are misrepresented, selectively framed, or detached from their proper context to suggest that vaccines are unsafe or ineffective[21]. The second is narrative evidence, in which anecdotes, testimonials, or individual experiences are used to raise concerns about vaccines and articulate skepticism in a more personal and emotionally resonant form[1,22,23]. These two forms of vaccine-skeptical content differ not only in their epistemic basis, but also in the challenges they pose for counterspeech.

Effectively responding to statistical-based vaccine-skeptical content requires a certain level of vaccine knowledge, statistical literacy, and access to credible sources. Counterspeakers must be able to identify where the evidence has been distorted and explain why the interpretation of data and statistics is misleading. Narrative-based vaccine-skeptical content poses a different challenge, as it derives its persuasive potency from the perceived authenticity of lived experience rather than average trends and generalizable information[24]. For this reason, countering a personal health experience story with statistics and numbers may appear dismissive and unacceptable, even when the statistical evidence itself is factually accurate. Instead, effective responses to narrative content may require acknowledging the experience behind the story while offering authentic counter-narratives that open discursive space for alternative interpretations, framings, or perspectives, allowing broader audiences to see beyond the singular case[23].

As AI technologies advance, conversational AI systems have increasingly been used to augment and collaborate with humans across a broad range of writing, learning, and training contexts[25-27]. In the field of counterspeech, however, much of the current landscape has embraced a more automation-oriented integration of AI by designing bots to directly confront objectionable content or publishing AI-generated counterspeech directly onto online platforms[28-31]. While such approaches promise scalability, they may obscure what makes counterspeech meaningful in the first place. Counterspeech is not merely a message that corrects, challenges, or refutes; it is also a manifestation of epistemic and discursive agency. When AI is used to replace human voices rather than to support them, it risks hollowing out the authenticity and agency that endows counterspeech with its moral, social, and civic power[13].

In this paper, we focus on an AI-assisted approach to augment human counterspeech writing in responses to statistical-based and narrative-based vaccine-skeptical content. The main goals are threefold: First, we examine whether AI-assisted writing can improve individuals' perceived effectiveness of their counterspeech messages while preserving authenticity. We further test whether these gains, in turn, increase counterspeakers' willingness to publicly post their counterspeech messages to exert influence on public discourse. Second, we investigate the characteristics that make counterspeech messages perceived as effective by ordinary users. Identifying these characteristics matters because it translates a vague goal of "writing better counterspeech" into concrete, designable message features that writing tools and training programs can be built to scaffold them. Third, we examine the added value of AI-assisted writing tools. Rather than treating AI assistance as a black box, we unpack what lay counterspeakers struggle with when writing on their own, what kinds of support they seek from AI, and whether these forms of AI support translate into greater message effectiveness.

To span a more realistic design space, our custom-built AI-assisted writing systems mirror how generative AI is commonly used in everyday writing tasks: 1) as a rewriting tool that refines an existing draft, or 2) as a co-writing tool that supports idea generation and composition from the outset. Within the two co-writing conditions, we further distinguish guided from unguided AI support. Unguided support resembles interactions with general-purpose chatbots such as ChatGPT, where users must formulate their own open-ended prompts to elicit help. Guided support, on the other hand, resembles the predefined function-based AI assistance embedded in writing tools such as Notion AI, which encode knowledge about message quality into the interface itself (Methods). We situate the study in a Reddit-like online environment, where vaccine-skeptical content appears as anonymous user posts and counterspeech takes the form of direct comments responding to those posts.

To evaluate our AI-assisted writing systems, we conducted a randomized controlled trial with 550 participants recruited from Prolific. We employed a between-subjects study design, where each participant was randomly assigned to one of the four conditions: guided AI co-writing, unguided AI co-writing, AI rewriting, and human-only. Each participant was asked to compose counterspeech messages in response to two vaccine-skeptical posts: one grounded in statistical evidence and one grounded in narrative evidence. A set of four posts (2 statistical-based and 2 narrative-based) was collected and adapted from a vaccine-related Reddit community to reflect the real-world vaccine-skeptical content that users may encounter online. To preserve the ecological validity of lay counterspeech production, we did not provide any prior training; participants were instructed only to submit a comment that they perceived as effective and persuasive in countering the skeptical content. This design allowed us to capture what effective counterspeech means from the perspective of ordinary users, and how they choose to use, adapt, or forgo AI assistance in pursuit of that writing goal. Fig.1 presents the experimental procedure and four experimental conditions (three AI-assisted writing + one human-only control).

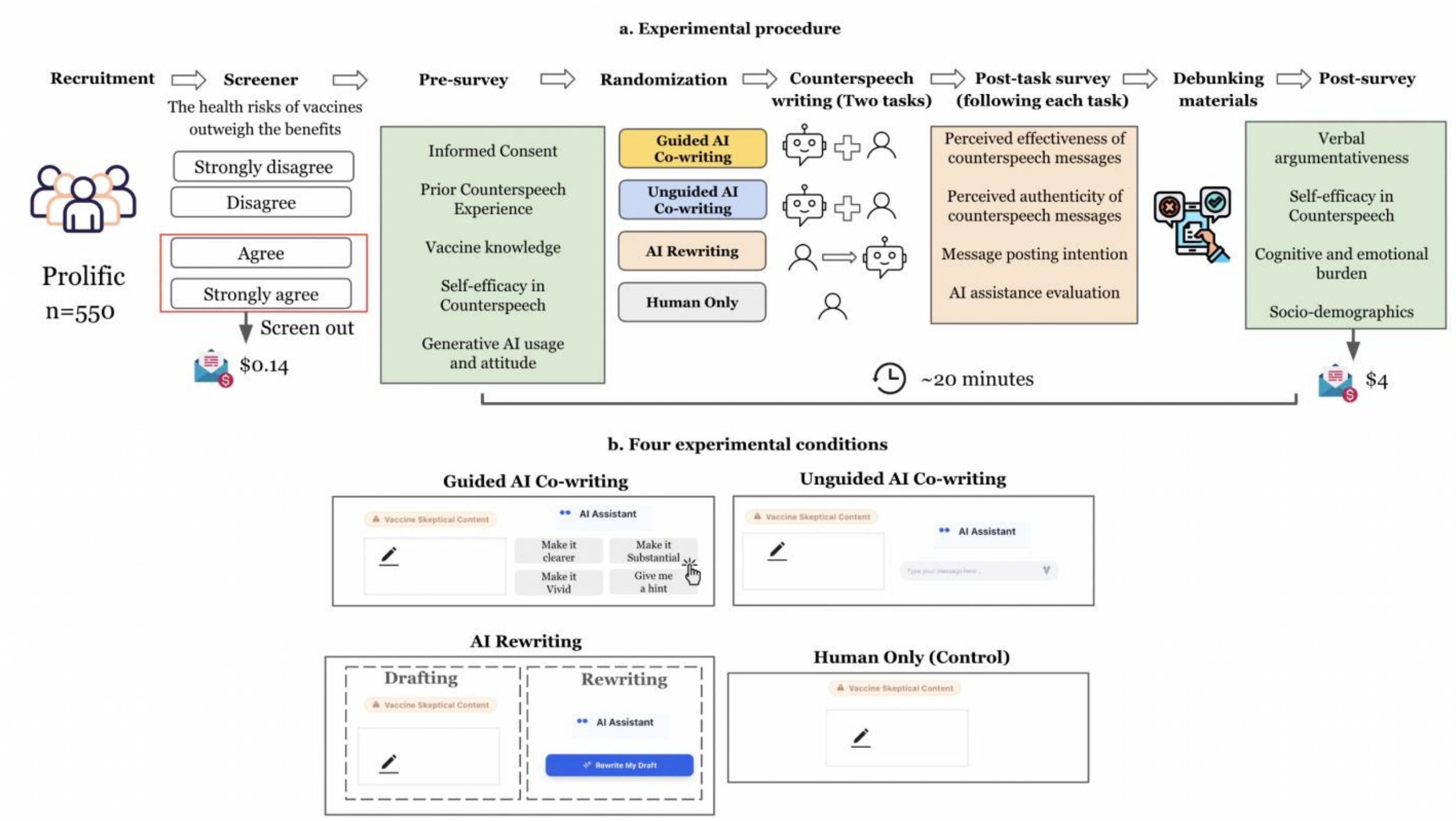

Figure 1. Visual illustrations of (a) experimental procedure and (b) four experimental conditions

## 2 Results

### 2.1 Effectiveness Gain versus Authenticity Cost in AI-assisted Counterspeech Writing

We conducted a series of liner mixed-effects models on the three main dependent variables – perceived message effectiveness, perceived authentic tone, and perceived authentic thoughts – with writing condition as a fixed effect and participant-level random intercepts to account for the repeated-measures structure, in which each participant contributed two data points (one for statistical post and one for narrative post). The analyses revealed statistically significant main effects of the writing condition on all three outcomes. Participants in all three AI-assisted conditions rated their counterspeech messages as more effective in countering vaccine-skeptical posts than participants in the human-only condition, as shown in Table 1. Fig.2 visualizes the differences in perceived message effectiveness and authenticity across the four writing conditions. At the same time, all three AI-assisted conditions impacted participants' perceptions of the messages' authenticity. AI assistance significantly reduced the perceived authentic tone. The effect on perceived authentic thoughts was more limited, as only participants in the unguided AI co-writing condition perceived their messages as significantly less reflective of their genuine thoughts than those in the human-only condition. To examine whether evidence type (i.e. statistical vs. narrative) moderated the effects of the AI-assisted writing condition on perceived message effectiveness and authenticity, we fitted models that included evidence type, writing condition, and their interaction, with random intercepts for participants. The results showed that the gains in perceived effectiveness and the reductions in authenticity associated with AI-assisted writing were largely consistent across both statistical and narrative posts, suggesting that evidence type plays a limited moderating role in shaping the main effects of AI-assisted writing condition on perceived effectiveness and authenticity (See detailed results in Supplementary Information S1).

**Table 1. Effects of AI-assisted writing condition on perceived message effectiveness and authenticity**

| | Perceived effectiveness | Perceived authentic tone | Perceived authentic thoughts |
|---|---|---|---|
| Guided AI co-writing | 6.07 (0.77)[a] | 5.77 (1.34)[a] | 6.24 (1.03)[ab] |
| Unguided AI co-writing | 5.98 (0.80)[a] | 5.83 (1.34)[a] | 5.91 (1.40)[b] |
| AI rewriting | 6.16 (0.76)[a] | 5.74 (1.36)[a] | 6.18 (1.22)[ab] |
| Human-only (control) | 5.69 (1.03)[b] | 6.28 (1.00)[b] | 6.35 (1.06)[a] |
| Omnibus ANOVA | $F(3, 510.97) = 9.99$, $p < 0.001$, marginal $R^2 = 0.045$ | $F(3, 513.19) = 7.33$, $p < 0.001$, marginal $R^2 = 0.03$ | $F(3, 511.11) = 4.20$, $p < 0.01$, marginal $R^2 = 0.02$ |
| Δ (Guided AI co-writing - Human-only) | Δ = 0.39, t(515.08) = 4.09, p < 0.001, d = 0.46 | Δ = -0.51, t(515) = -3.73, p < 0.001, d = -0.400 | Δ = -0.12, t(515.10) = -0.90, p =0.68, d = -0.10 |
| Δ (Unguided AI co-writing - Human-only) | Δ = 0.29, t(515.09) = 3.06, p < 0.001, d = 0.34 | Δ = -0.46, t(515.02) = -3.39, p < 0.001, d = -0.36 | Δ = -0.45, t(515.11) = -3.44, p <0.01, d = -0.38 |
| Δ (AI rewriting - Human-only) | Δ = 0.48, t(515.17) = 5.17, p < 0.001, d = 0.56 | Δ = -0.54, t(515.12) = -4.16, p <0.001, d = -0.43 | Δ = -0.17, t(515.20) = -1.35, p = 0.39, d = -0.14 |

Note: All three outcomes were measured on 7-point Likert scales. Values in the upper rows are means with standard deviations in parentheses. Within each column, means that do not share a subscript differ significantly at P < .05 based on Tukey-adjusted pairwise comparisons. Contrasts were estimated from mixed-effects models with participant-level random intercepts, in which each participant contributed two data points. Omnibus F-tests use Satterthwaite-approximated degrees of freedom, and effect sizes are reported as marginal $R^2$. Δ represents the difference between each AI-assisted condition and the human-only control, computed by subtracting the control mean from the AI-assisted condition mean; positive values indicate higher ratings in the AI-assisted conditions against the control, with Dunnett-adjusted, two-sided P values. n of observations =1,022.

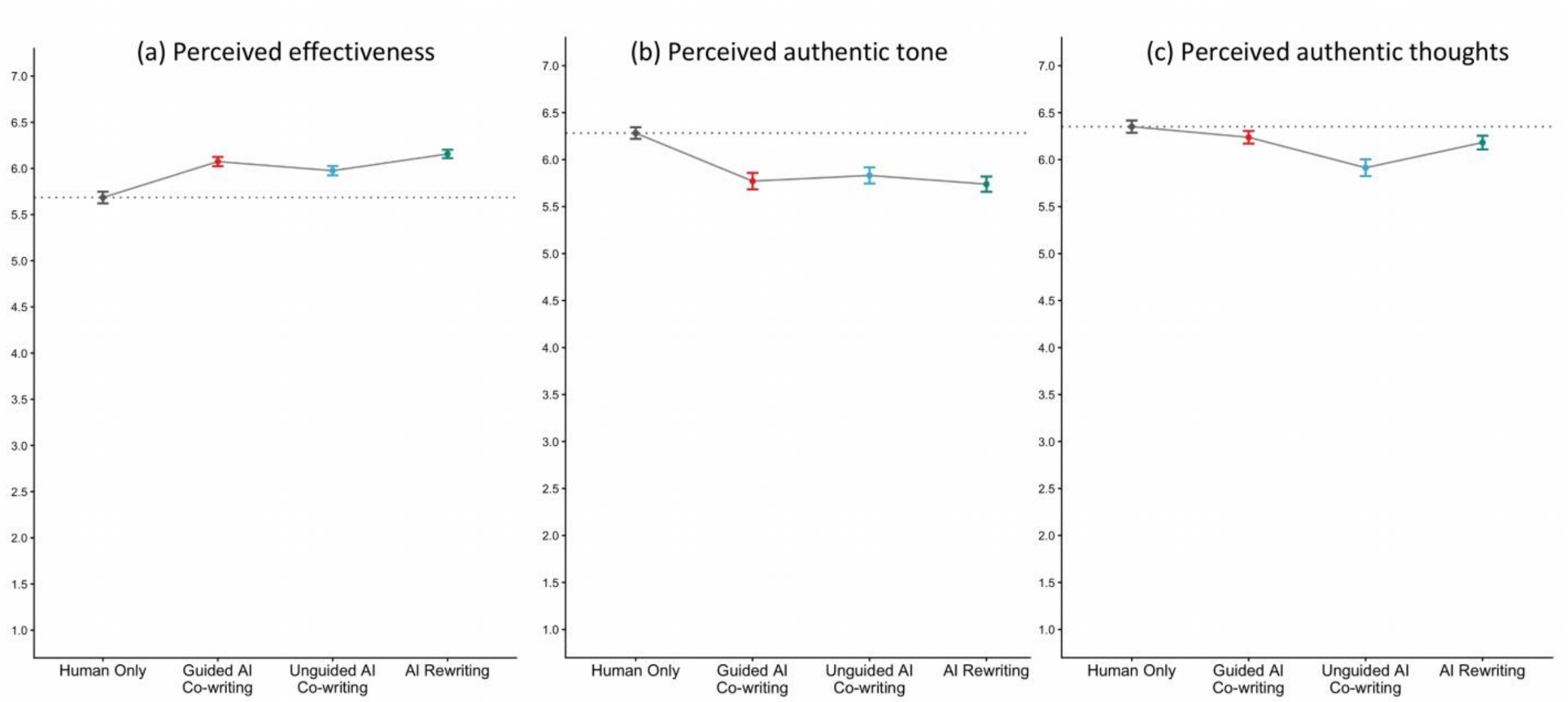

**Figure 2. Perceived message effectiveness and authenticity across four writing conditions**. (a) perceived message effectiveness; (b) perceived authentic tone; (c) perceived authentic thoughts

## 2.2 Factors predicting counterspeech posting intention

A counterspeech message can only fulfill its discursive function if its author is willing to post it publicly. To examine what drives counterspeech posting intention, we first tested whether AI-assisted writing condition or evidence type influenced willingness to post using a linear mixed-effects model with AI-assisted writing condition, evidence type, and their interaction as fixed effects and random intercepts for participants. Evidence type showed a significant main effect, $F(1, 500.83) = 19.08$, $p<0.001$: participants were more willing to post counterspeech written in response to narrative than statistical posts, $\Delta = 0.25$, 95% CI [0.14, 0.36], $p <0.001$. Neither the main effect of the AI-assisted writing condition, $F(3, 511.83)= 0.86$, $p =0.462$, nor the condition X evidence type interaction, $F(3, 500.84) = 0.18$, $p =0.907$, reached significance.

We next examined whether perceived message characteristics and individual differences predicted counterspeech posting intention. Separate linear regression models were fitted for statistical and narrative posts, with posting intention as the outcome and three categories of predictors: perceived message characteristics (perceived effectiveness, perceived authentic tone, and perceived authentic thoughts), individual counterspeech characteristics (verbal argumentativeness and prior counterspeech experience), and demographic characteristics (age, gender, race, income, education, and political orientation).

Across both post types (Table 2), perceived message effectiveness was the strongest predictor of posting intention (statistical posts: $\beta=0.36$, 95% CI = [0.29, 0.43], $p <0.001$; narrative posts: $\beta=0.28$, 95% CI = [0.19, 0.37], $p <0.001$). Perceived authentic thoughts also positively predicted posting intention (statistical posts: $\beta=0.10$, 95% CI = [0.01, 0.19], $p = 0.027$; narrative posts: $\beta=0.11$, 95% CI = [0.01, 0.21], $p =0.025$), whereas perceived authentic tone did not (statistical posts: $\beta=0.04$, 95% CI = [-0.04, 0.13], $p =0.322$; narrative posts: $\beta=0.06$, 95% CI = [-0.05, 0.16], $p =0.277$).

At the individual level, participants higher in verbal argumentativeness (statistical posts: $\beta= 0.14$, 95% CI = [0.06, 0.22], $p <0.001$; narrative posts: $\beta=0.18$, 95% CI = [0.09, 0.26], $p <0.001$) and with greater prior counterspeech experiences (statistical posts: $\beta=0.26$, 95% CI = [0.18, 0.33], $p <0.001$; narrative posts: $\beta=0.27$, 95% CI= [0.20, 0.35], $p <0.001$) reported stronger posting intentions. No demographic variable was a significant predictor (all Ps >0.05).

**Table 2. Regression analysis of counterspeech posting intention on message-level and individual-level characteristics.**

| Predictor | Statistical Post | | Narrative Post | |
|---|---|---|---|---|
| | Estimate | 95% CI | Estimate | 95% CI |
| Intercept | 0.03 (0.11) | -0.19, 0.25 | 0.14 | -0.08, 0.37 |
| Perceived message characteristics | | | | |
| Perceived effectiveness | 0.36 (0.04)*** | 0.29, 0.43 | 0.28 (0.05)*** | 0.19, 0.37 |
| Perceived authentic thoughts | 0.10 (0.05)* | 0.01, 0.19 | 0.11 (0.05)* | 0.01, 0.21 |
| Perceived authentic tone | 0.04 (0.04) | -0.04, 0.13 | 0.06 (0.05) | -0.05, 0.16 |
| Individual counterspeech characteristics | | | | |
| Prior counterspeech experience | 0.26 (0.04)*** | 0.18, 0.33 | 0.27 (0.04)*** | 0.20, 0.35 |
| Verbal argumentativeness | 0.14 (0.04)*** | 0.06, 0.22 | 0.18 (0.04)*** | 0.09, 0.26 |
| Individual demographic characteristics | | | | |
| Age | 0.03 (0.04) | -0.04, 0.11 | 0.01 (0.04) | -0.06, 0.09 |
| Gender [Male] | 0.02 (0.08) | -0.12, 0.17 | -0.05 (0.08) | -0.20, 0.11 |
| Gender [Third] | 0.09 (0.25) | -0.40, 0.57 | 0.42 (0.26) | -0.09, 0.93 |
| Race [Black] | -0.01 (0.11) | -0.23, 0.21 | 0.07 (0.12) | -0.16, 0.30 |
| Race [Asian] | 0.02 (0.15) | -0.28, 0.32 | 0.07 (0.16) | -0.25, 0.40 |
| Race [Other] | -0.19 (0.14) | -0.47, 0.09 | -0.24 (0.15) | -0.54, 0.05 |
| Income [$60k–$99k] | 0.00 (0.09) | -0.18, 0.18 | 0.02 (0.10) | -0.17, 0.21 |
| Income [≥$100k] | 0.03 (0.09) | -0.14, 0.20 | 0.01 (0.09) | -0.17, 0.19 |
| Education [Some college] | -0.00 (0.13) | -0.25, 0.25 | -0.05 (0.13) | -0.31, 0.22 |
| Education [College or higher] | -0.08 (0.11) | -0.30, 0.14 | -0.16 (0.12) | -0.38, 0.07 |
| Political Orientation | 0.01 (0.04) | -0.06, 0.08 | 0.03 (0.04) | -0.04, 0.11 |
| $R^2$ | 0.404 | | 0.311 | |
| Adj-$R^2$ | 0.384 | | 0.288 | |

Note: Coefficients are standardized beta (β), with SE reported in the parentheses. All continuous predictors and the outcome were z-standardized on the full sample prior to model fitting; categorical predictors are dummy-coded, with coefficients representing differences from the reference group. Reference groups are women (gender), White (race), below $60,000 (income), and high school or less (education). Separate models were fitted for statistical and narrative posts. Tests are two-sided. *P<0.05; ***P<0.001. n = 509 for statistical-based posts; n = 513 for narrative-based posts.

To quantify the contribution of each predictor, we used the relaimpo package in R to calculate relative importance metrics for two reduced models containing the three message characteristics and two individual counterspeech characteristics. Demographic variables were excluded given their non-significance in the full models. For statistical posts (Fig.3), the five predictors jointly explained 39.9% of the variance in posting intention, with perceived effectiveness accounting for the largest share of the explained variance (44.6%), followed by prior counterspeech experience (20.8%), verbal argumentativeness (14.4%), perceived authentic thoughts (11.3%) and perceived authentic tone (8.9%). For narrative posts, the predictors explained less variance overall (29.6%), with prior counterspeech experience contributing the largest share of the explained variance (32.6%), followed by perceived effectiveness (29.9%), verbal argumentativeness (20.0%), perceived authentic thoughts (10.0%) and perceived authentic tone (7.6%).

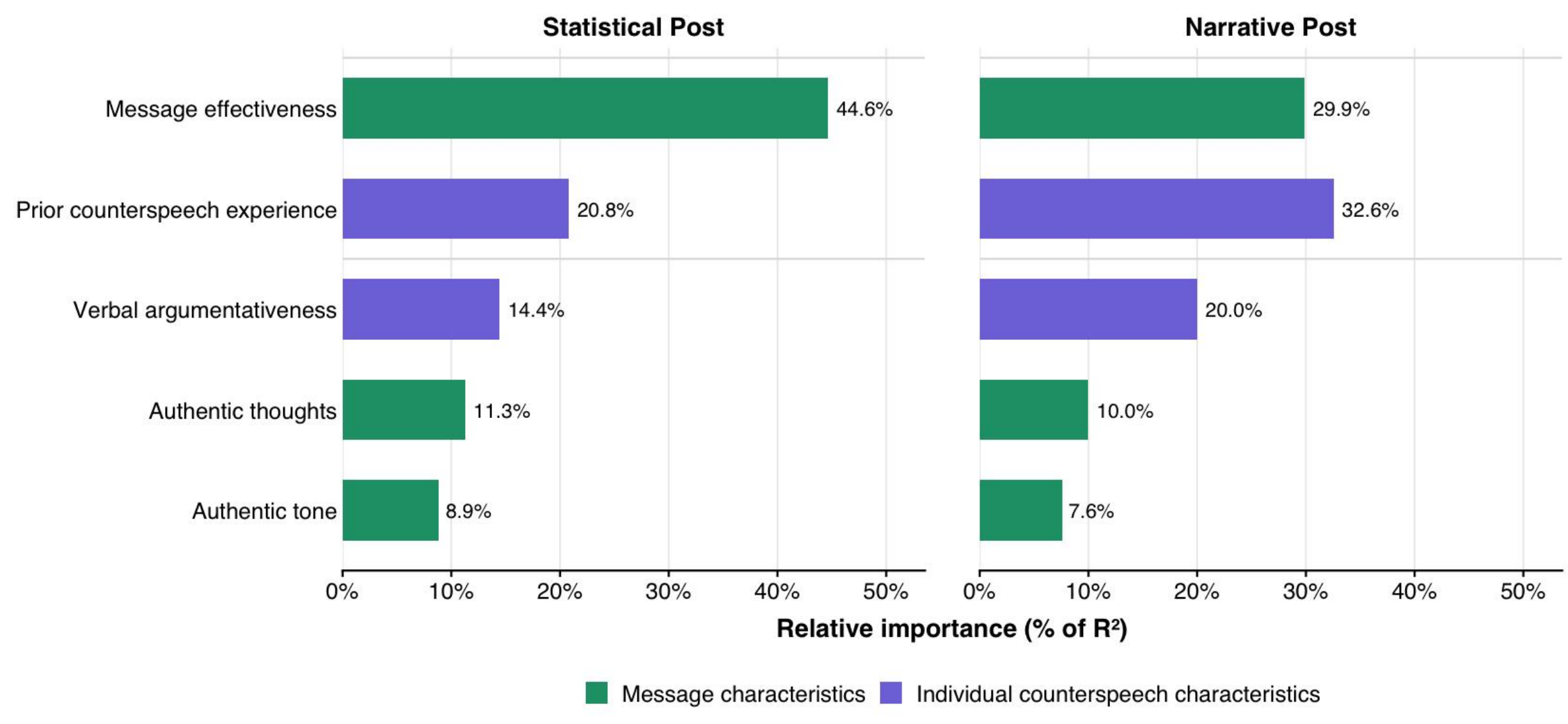

**Figure 3. Relative importance of factors predicting counterspeech posting intention, by evidence type**

These results show that while AI assistance does create a tradeoff – increasing perceived effectiveness while reducing authentic tone – these two outcomes matter unequally for whether counterspeech reaches the public: perceived effectiveness was among the strongest drivers of posting intention in response to both types of vaccine-skeptical posts, whereas authentic tone did not significantly predict it. The dimension AI assistance enhances is thus precisely the one that most strongly propels posting, while the dimension it erodes contributes little.

## 2.3 What characteristics make counterspeech messages perceived as effective?

Our analysis thus far has shown that AI-assisted writing improved the perceived effectiveness of the messages lay counterspeakers crafted, and that perceived effectiveness was in turn a key driver of their intention to post these messages publicly. These findings raise a further question: which message characteristics underlie perceived counterspeech effectiveness, and how, if at all, does AI assistance shape them? To address this, we leveraged the draft-submission message pairs from the AI rewriting condition, which allow a within-person comparison of messages written by the same participant before and after AI assistance.

We begin by asking whether AI rewriting had an effect on the evidentiary form users deployed in their counterspeech messages. In total, 582 messages (291 draft-submission pairs) from 146 participants were manually coded for their evidentiary forms. Our results showed that across message pairs, the overwhelming majority of counterspeech messages (85% to 94.5%) relied on logical argumentation or statistical evidence, regardless of the evidentiary form of the original post. This dominance was equally present in initial drafts and final submissions. Participants thus appeared to approach counterspeech as a task of reasoning, logical argumentation, or evidence-based correction from the outset, and AI rewriting preserved, rather than reshaped, this strategic orientation. Because participants in this condition knew AI assistance would be available, their initial drafts might not reflect how people write counterspeech unaided. To rule this out, we also manually coded the counterspeech messages submitted in the human-only condition (n=270), where no AI assistance was offered or expected. The results were consistent: nearly 90% of messages in human-only condition likewise relied on logical argumentation or statistical evidence, regardless of the evidence used in the original post.

During manual coding of the draft-submission pairs, we observed that final submissions were often substantially more elaborate than initial drafts, even when the underlying evidentiary logic remained unchanged. To quantify this behavioral pattern, we use the term message elaborateness to refer to the degree to which a message has been linguistically and cognitively developed beyond its minimal expression. We operationalize this construct along two complementary dimensions: depth of development, captured by analytical thinking and information density, which index how much reasoning and substantive content a message packs in; and extent of development, captured by text complexity and message length, which index the message's surface-level linguistic and structural development (see Methods). We then compared these four message elaborateness features across three message groups: human-written messages from the human-only condition, initial drafts from the AI rewriting condition, and final submissions from the AI rewriting condition. This three-group comparison serves two purposes: first, it tests whether participants' initial drafts in AI rewriting condition (rewrite draft) were comparable to unaided human writing (human-only submission) on message elaborateness features; second, it tests whether the final, AI-assisted submissions (rewrite submission) exceeded that human-writing baseline (rewrite draft and human-only submission).

As shown in Fig.4, AI rewritten submissions exhibited significantly greater analytical thinking and information density, and were longer and more complex, than both submissions in human-only condition and the original drafts (but unsubmitted) in AI rewriting condition (all Ps<0.05; see Supplementary Information S2). By contrast, human-only final submissions and those original drafts before applying AI rewriting did not differ significantly on any message elaborateness features (all Ps>0.05; see Supplementary Information S2).

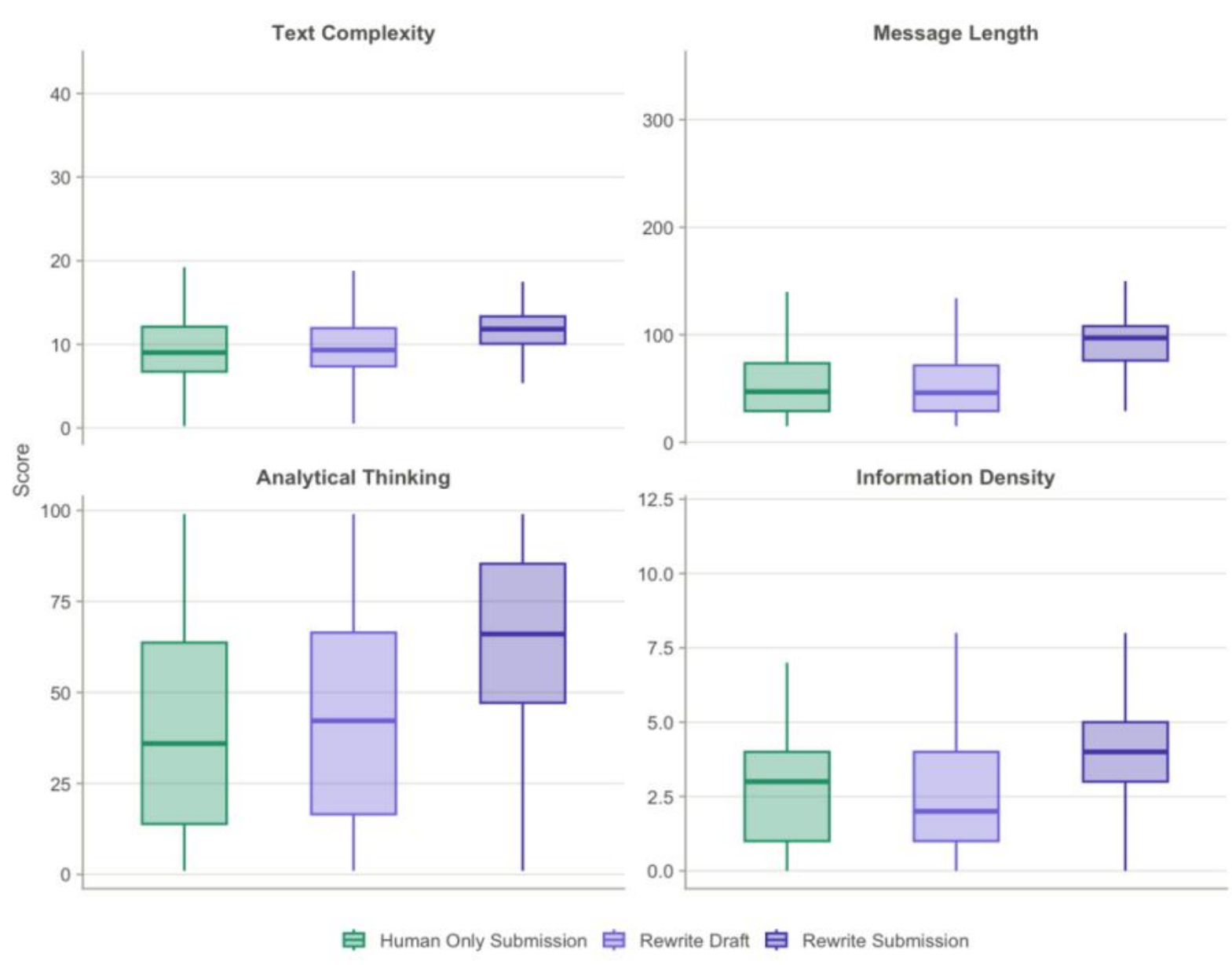

**Fig.4. Group comparisons of four message elaborateness features between human-only final submission, rewrite draft, and rewrite final submission**

These findings suggest two important aspects of how users deployed AI in crafting counterspeech messages. First, the increased message elaborateness in the AI-rewritten final submissions did not appear to reflect a difference in how users approached the task. Instead, message elaborateness was added during the AI rewriting process itself. Second, users appeared to value precisely this elaborate expression, choosing to use AI to refine less elaborate drafts before submitting more elaborate final versions. AI assistance, in other words, helped participants transform original drafts comparable to unaided human writing into longer, more complex, analytically developed, and information-rich counterspeech messages.

We next ask whether this observed behavioral pattern is in line with users' own evaluation of their messages. We computed the four elaborateness measures for all final submitted messages (n=1,022) and fitted separate linear regression models for statistical (n=509) and narrative posts (n=513) separately, with the standardized elaborateness features as predictors and standardized perceived effectiveness as the outcome. Consistent with the behavioral inference above, we find that across both post types, message length (statistical posts: β=0.14, 95% CI = [0.01, 0.27], p =0.03; narrative posts: β=0.12, 95% CI = [0.02, 0.21], p =0.015), text complexity (statistical posts: β=0.11, 95% CI = [0.02, 0.21], p =0.016; narrative posts: β=0.10, 95% CI = [0.02, 0.19], p =0.013), and information density (statistical posts: β=0.15, 95% CI = [0.04, 0.27], p =0.011; narrative posts: β=0.11, 95% CI = [0.00, 0.21], p =0.04) were positively associated with perceived effectiveness. Analytical thinking was significantly associated with perceived effectiveness only for counterspeech responding to narrative posts, β=0.13, 95% CI = [0.05, 0.21], p =0.002, but not statistical posts, β=0.10, 95% CI = [-0.00, 0.20], p =0.061.

This pattern supports the proposed pathway in which message elaborateness features contributed to users' perceived effectiveness of their counterspeech messages. Given that perceived effectiveness was also positively associated with willingness to post, we further examined whether perceived effectiveness mediated the relationship between message elaborateness features and counterspeech posting intention. Using the lavaan package in R, we estimated separate mediation models for counterspeech responding to statistical and narrative posts. In each model, the four message elaborateness features were entered as predictors, perceived effectiveness was specified as the mediator, and willingness to post was specified as the outcome. All variables were standardized, and AI-assisted writing condition was included as a covariate. Standard errors were estimated using 5,000 bootstrap draws.

Results revealed that perceived effectiveness partially explains why more elaborate counterspeech may increase willingness to post, but the pattern differs by elaborateness feature. For narrative posts, text complexity, β=0.04, SE=0.02, 95% CI= [0.00, 0.08], p =0.037, information density, β=0.05, SE=0.02, 95% CI= [0.01, 0.09], p =0.03, and analytical thinking, β=0.04, SE=0.02, 95% CI= [0.00, 0.08], p =0.033, all had a significant positive indirect effect on willingness to post through perceived effectiveness. The indirect effect of message length was also positive but marginally significant, β=0.04, SE=0.02, 95% CI = [0.00, 0.08], p =0.07. For statistical posts, the indirect effects through perceived effectiveness remained significant for text complexity, β=0.05, SE=0.02, 95% CI = [0.01, 0.10], p =0.013, and information density, β=0.08, SE=0.03, 95% CI= [0.03, 0.13], p=0.004. The indirect effect of message length was marginally significant, β=0.06, SE=0.03, 95% CI =[-0.00, 0.11], p =0.064, whereas the indirect effect of analytical thinking was not significant, β=0.03, SE=0.03, 95% CI= [-0.03, 0.08], p =0.28. Supplementary Information S3 provides detailed results for mediation analyses.

### 2.4 AI support translates users' elaboration needs into more elaborate forms of counterspeech

Having observed that AI support helped individuals write more elaborate messages, which they perceived as more effective and tended to favor for publicly posting, we next ask whether elaboration lies at the heart of users' writing needs during counterspeech composition. By elaboration, we mean the process of developing a more elaborate message. Specifically, we take a closer look at the elaboration-oriented assistance participants sought from the AI in the guided AI co-writing condition. We focused on this specific condition because it is the only one in which the AI interface included functions that directly supported message elaborateness. We designed two functions that served different but related purposes: “Give me a hint” can be used throughout the writing process, helping participants get started or identify relevant facts and arguments as they wrote, whereas “make it substantial” could only be used with an existing draft and was designed to expand and substantiate that draft. These guided AI functions captured two common elaboration needs in crafting persuasive messages: finding something meaningful to say and developing it into a fuller form. This condition therefore allows us to assess (1) to what extent participants voluntarily used these two elaboration-oriented functions during the writing process, reflecting their motivation to produce more elaborate messages, and (2) whether using these elaboration-oriented functions actually yielded more elaborate messages than using the other two functions in this condition, “make it vivid” and “make it clearer”, which were designed to improve expression rather than expand on message content.

To unpack how participants in the guided AI co-writing condition used the elaboration-oriented functions, we traced each participant’s writing path from their first action to final submission. Because participants could use any AI function multiple times before submitting their final counterspeech, we recorded each intermediate message state during the writing process. Specifically, whenever a participant triggered an AI function and then updated the message in the comment box, a new message version was created. This procedure yielded 926 message versions across 220 chat sessions. At the most basic level, writing paths in this condition fall into two broad types: users started by requesting a hint or started by writing some text. If they start with writing some text, then the other functions (“make it…”) become available.

A first, simple test of users’ need for elaboration can be observed in how often they begin by asking for a hint. Slightly over half of the writing sessions began with the hint chat (113/220, 51%); hint-first sessions were more common for statistical posts (59%) than for narrative posts (41%). Furthermore, although the remaining sessions (107/220, 49%) began with participants drafting on their own, 72% of these (77/107) then used the elaboration-oriented functions – “make it substantial” or “Give me a hint” – before submitting their counterspeech. In total, 86% (190/220) of writing sessions involved at least one use of an elaboration-oriented function, suggesting that elaborate writing constitutes a major need for ordinary users crafting counterspeech with guided AI support.

We then compared the four message elaborateness features between chat sessions that used elaboration-oriented functions (n=190) and those that did not (n=30), using two-sided Welch’s t-tests; P values were adjusted for multiple comparisons across the four features using the Benjamini-Hochberg procedure (Fig.5).

Messages from sessions that used any elaboration-oriented functions were longer (M=78.38, SD=34.83) than those from sessions that did not (M=63.59, SD=23.32; t(49.5)=2.95, p <0.01, Cohen's d=0.50, 95% CI [0.18, 0.83]), higher in text complexity (M=11.87, SD=4.09 vs. M=9.55, SD=2.29; t(59.75)=4.46, p <0.001, Cohen's d=0.70, 95% CI [0.34, 0.98]), and higher in information density (M=3.33, SD=2.00 vs. M=2.52, SD=1.62; t(42.26)=2.44, p =0.03, Cohen's d=0.45, 95% CI [0.09, 0.79]). The two groups did not differ in analytical thinking (M=62.21, SD=25.08; vs. M=60.08, SD=28.19; t(35.1)=0.38, p =0.70, Cohen's d=0.08, 95% CI [-0.32, 0.50]). Collectively, these results suggest that lay counterspeakers overwhelmingly and voluntarily sought elaborate writing support from AI, and that support measurably increased the elaborateness of their messages.

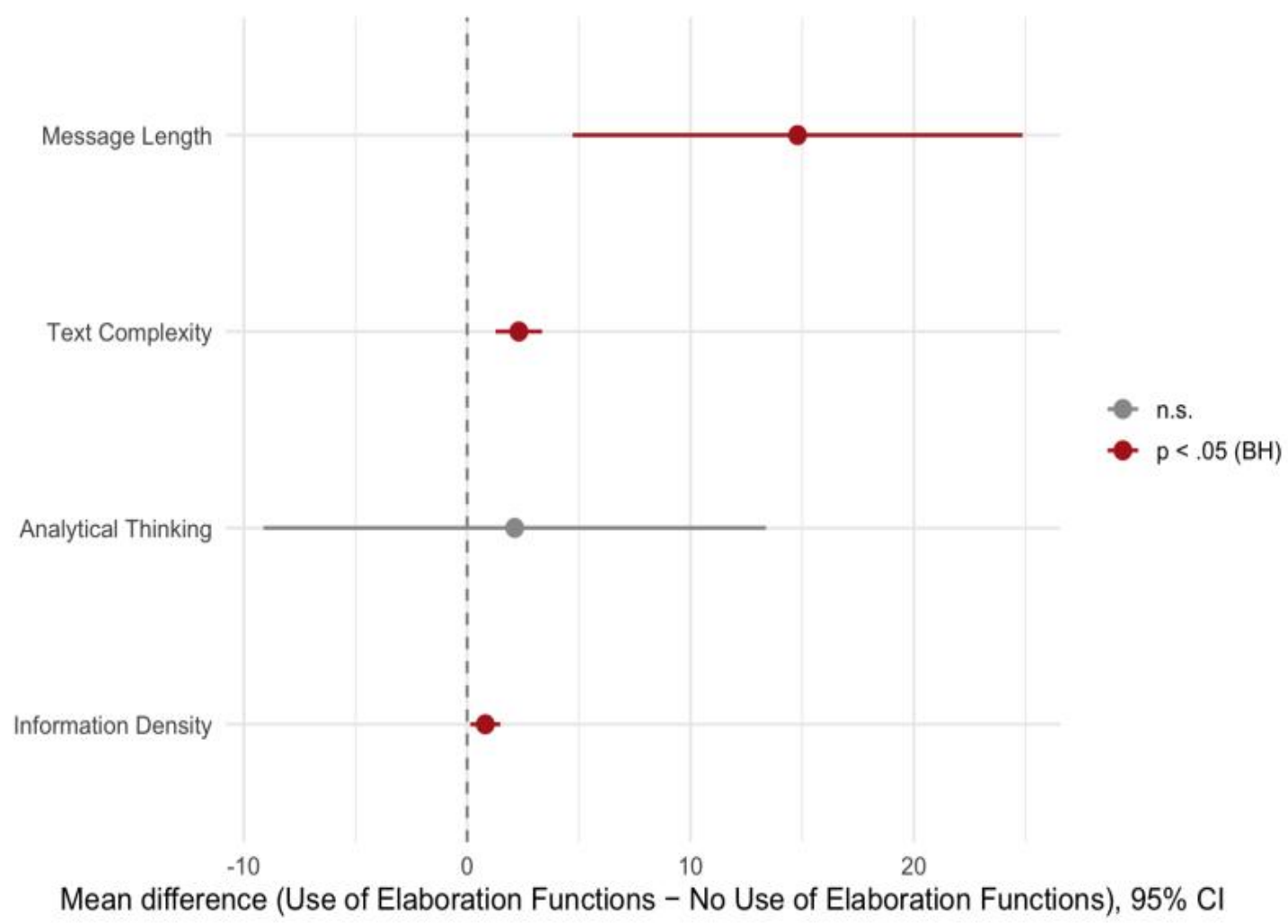

**Fig.5. Mean differences in four message elaborateness features between messages using versus not using elaboration-oriented AI functions (guided AI co-writing condition)**

### 2.5 Will message elaborateness contributes to evaluation of AI assistance?

The observed patterns of message elaborateness suggest that participants sought and benefited from AI assistance in crafting more elaborate counterspeech messages. However, it remains unclear whether participants recognized and attributed these improvements to AI, evaluating it as more useful, easier to use, or contributing to the persuasiveness of their messages. To examine this association, we fitted multiple regression models to test the relationships between the four elaborateness features and three dimensions of AI evaluation: perceived usefulness, ease of use, and perceived AI contribution to message persuasiveness. Results showed that, across all three dimensions, analytical thinking and message length were significantly positively associated with all three outcomes. These findings suggest that participants' favorable assessments of AI assistance were partly linked to the elaborative qualities of the messages they produced with AI support. More detailed results are reported in Supplementary Information S4.

## 3 Discussion

Community-driven counterspeech represents a promising approach to curbing widespread vaccine skepticism and engaging laypeople in more informed civic participation around controversial scientific and societal topics. Yet this form of participation places substantial demands on ordinary users who will often lack the skill, knowledge, or rhetorical resources needed to respond effectively in countering objectionable content such as vaccine misinformation and skeptical content. Our research asks whether AI can help address this challenge through assisting people in crafting counterspeech messages against statistical-based and narrative-based vaccine-skeptical content using three custom-built, generative AI-powered counterspeech writing systems.

Results indicate that AI can help laypeople produce counterspeech messages perceived as more effective, regardless of the type of the skeptical post, the stage at which AI is involved in the writing process or whether the AI support is guided. This approach captures the distinct value of AI – the ability to increase elaborate writing that allows counterspeech messages to more fully and persuasively address misleading content. Specifically, our analysis revealed that users (1) sought elaboration-oriented support from AI; (2) preferred to submit more elaborate messages assisted by AI; (3) felt that more elaborate messages were more effective; and 4) felt that, because they were more effective, elaborate messages were more worthy of posting. In contrast, unaided participants faced greater difficulty writing counterspeech messages on their own, experiencing higher emotional and cognitive burdens (Supplementary Information S5), resulting in messages they perceive to be lower in quality.

While our study focused on counterspeakers' behaviors, perceptions, and evaluations of their own messages, its central finding that elaborateness features constitute a key ingredient of perceived message effectiveness is consistent with a long tradition in persuasion research. In line with the elaboration likelihood model[32], prior work has identified message length, information richness, and argumentative elaboration as important message characteristics that can enhance persuasive potency[33-35]. Recent work in political persuasion further supports AI's distinctive strength in elaboration support, as information density emerges as a key mechanism underlying AI persuasiveness[36].

These findings show that ordinary users' intuitions that effective counterspeech is more elaborate are accurate; if they want to be effective counterspeakers, they should seek to provide elaborate messages, and also, they are likely to lack the ability to do so on their own. In this context, AI tools serve as a critical bridge between users' motivation and capability by facilitating more elaborate writing.

More importantly, our study found that perceived improvements in message quality were associated with lay users' greater willingness to publicly counter-speak. Across the message-level and individual-level factors associated with counterspeech posting intention, perceived message effectiveness was the strongest predictor of willingness to respond to statistical-skeptical posts, and the second strongest predictor for countering narrative-skeptical posts. This aligns with prior research showing that ordinary social media users are less likely to engage with problematic content when they doubt the impact of their messages[16,17]. Our research thus suggests that AI assistance offers a concrete lever for elevating behavioral intention by addressing the efficacy doubts that have previously kept lay users on the sidelines. This message-level boost was also more important than – or at least comparably important to – individual trait characteristics (e.g., verbal argumentativeness) and experiential characteristics (e.g., prior counterspeech experience) in driving counterspeech engagement, pointing to a practical pathway for bystanders to step into counterspeech roles even without extensive practice or a naturally argumentative disposition.

In the context of counterspeech writing, preserving humans' authentic expression is also important. One of the features that gives counterspeech its persuasive power is the fact that it signals the thoughts and feelings of a real person, not an algorithm or bot. In our study, although AI involvement in counterspeech writing understandably reduced the perceived authenticity of tone, it largely preserved the authentic thoughts that counterspeakers sought to convey, especially when writers were guided with AI functions or used AI primarily for post-hoc rewriting. This pattern aligns with prior work suggesting that greater user control over AI-assisted writing can help preserve user agency[37]. Moreover, only perceived authentic thoughts, but not tone, significantly predicted counterspeech posting intention. This suggests that what ultimately matters to counterspeakers is whether the message is effective and faithfully conveys their intended meanings, rather than whether its tone is preserved intact in AI-assisted rewriting.

What are the implications of these findings? We argue they suggest value in a level-up pathway for AI-assisted counterspeech – one in which humans are not replaced by AI but use AI to enhance their own efforts at community-driven moderation. This approach captures the distinct value of AI – the ability to increase elaborate writing that allows counterspeech messages to more fully and persuasively address misleading content. At the same time, it preserves the key ingredient that differentiates counterspeech from top-down approaches, the voluntary agency and accountability of individuals who choose to speak up. With AI enhancement, "leveled up" participants' voluntary uptake of elaboration support reflects a naturalistic form of human-AI collaboration, in which users selectively engage AI assistance when it aligns with their communicative goals. This suggests that AI-facilitated counterspeech holds potential to take root organically in real-world online contexts. In an online environment increasingly shaped by brevity and emotionally arousing content, our findings align with recent work[36] in showing that laypeople still value reasoned and elaborate communication. By supporting users in developing richer arguments and more substantive responses, AI may contribute to more meaningful human-to-human exchanges rather than displacing the human voice from public discourse.

A key scope condition of our research is that perceptions of message effectiveness come from counterspeakers themselves, rather than the audiences who would receive these messages. Although gaps between authors' and audiences' perceptions of message effectiveness are not uncommon[38,39], our study does not address whether AI-assisted counterspeech generates actual persuasion outcomes on audiences. Instead, it asks whether AI can help ordinary users feel that their responses are sufficiently effective, authentic, and worth posting, thereby enabling them to enter contested online spaces and voice their positions. Future research should examine how counterspeakers' perceived effectiveness maps onto audience evaluations and downstream engagement with AI-assisted counterspeech messages.

Another scope condition is that we only evaluated the perceived efficacy of AI-assisted counterspeech in the context of vaccine-skeptical content. Within this context, elaborate writing emerged as a key pathway through which ordinary users' writing needs and AI's strengths converge. While this elaboration mechanism aligns with prior persuasion scholarship, it remains unclear whether this pattern will extend to counterspeech in other arenas, such as hate speech or harassment, where alternative message characteristics, such as empathy or moral framing may assume greater prominence[28,40]. Future research should examine different deployment contexts, as well as AI functions and features, to identify the boundary conditions governing the efficacy of AI-assisted writing in counterspeech.

## 4 Materials and Methods

### 4.1 Experimental Procedure

The study was approved by the Institutional Review Board at a research-based university as exempt research (protocol no. IRB0150324). All participants were recruited from Prolific and were compensated $4.00 for completing the study. To minimize the influence of preexisting vaccine attitudes on task completion, we included a screener item ("The health risks of vaccines outweigh the benefits"[41]) and excluded participants who endorsed it (selecting "Agree" or "Strongly agree"). Screened-out participants received $0.14 in compensation. Eligible participants proceeded to the informed consent page on Qualtrics, in line with the IRB's recommendation for this exemption category. Those who consented to participate then completed a pre-study survey, which included questions about their vaccine knowledge, prior counterspeech experience, self-efficacy in writing counterspeech, and their use of and attitudes toward generative AI. Participants were subsequently randomly assigned to one of four conditions on our custom-built AI-assisted counterspeech writing system developed using Google AI Studio. A guide tour tailored to each condition was provided to help participants understand the interface and their task.

On the experiment platform, participants read two anonymous user posts containing skeptical claims about COVID-19 vaccine efficacy and were asked to write and submit a comment countering each post. The two posts differed in evidentiary form: one featured statistical evidence, while the other featured narrative evidence. Posts were presented one at a time, in randomized order. To reduce stimulus-specific effects, we drew from a pool of two statistical and two narrative posts, collected and adapted from r/DebateVaccines (see Supplementary Information S6).

In the AI-assisted conditions, an AI assistant was available to support counterspeech writing in the manner specified by each condition. Participants could use the assistant as often as they wished, or not at all, with the goal of crafting a comment that they perceived persuasive and effective in countering the vaccine-skeptical post. Participants in the control condition completed the task independently, without AI assistance. The study employed a between-subjects design comparing three AI-assisted conditions with a human-only control condition. The AI-assisted conditions varied along two dimensions: the writing stage at which assistance was provided (co-writing during drafting vs. rewriting after drafting) and the assistance mode (guided through predefined writing functions vs. unguided through open-ended user prompts).

In the guided AI co-writing condition, participants had access to four predefined AI functions during drafting (“Give me a hint”; “Make it substantial”, “Make it vivid”, and “Make it clear”), which provided structured writing assistance similar to domain-specific AI-assisted writing systems such as Notion AI. In the unguided AI co-writing condition, participants interacted with the assistant through open-ended prompts – mirroring generic use of generative AI – to receive assistance during drafting. In the AI rewriting condition, assistance became available only after participants had composed an initial draft on their own, allowing them to refine and polish it (Fig.1).

The AI assistant generated responses in real time using Google’s Gemini language model (gemini-3-flash-preview). To make the assistance responsive to participants’ ongoing writing, the assistant received both the vaccine-skeptical post being addressed and the participants’ current draft from the comment box, allowing it to take the post context and the participants’ progress into account whenever assistance was requested. Given the limited size of the chat window, the assistant was instructed to keep each response concise. We present the system prompts used for each AI-assisted condition in the Supplementary Information (S7).

Once satisfied with their comments, participants submitted their counterspeech messages and completed a post-task survey, which assessed the perceived effectiveness and authenticity of their submitted messages, their willingness to post the responses in public venues, and, in the AI-assisted conditions, their evaluation of AI assistance. After the writing tasks, participants were shown debunking materials tailored to the posts they had encountered and were then redirected to Qualtrics to complete a set of post-study questions.

## 4.2 Participants

Based on a priori power analysis assuming a small effect size (f = 0.15), a significance level of α =0.05, and 80% power for detecting main and interaction effects in a four-condition between-subjects ANOVA; conducted in G*Power version 3.1, a minimum sample of 492 participants was required. Our final sample included 550 participants who completed all study procedures: 134 in the guided AI co-writing condition, 135 in the unguided AI co-writing condition, 146 in the AI rewriting condition, and 135 in the human-only condition. Randomization was maintained through the Qualtrics randomizer. The slightly larger sample in the AI rewriting condition reflects the logic of the “evenly present elements” setting: Qualtrics balances condition presentation when participants enter the randomizer, rather than at survey completion. Therefore, participants who abandoned or re-entered the survey after assignment could affect Qualtric’s internal counts without contributing valid completed responses, producing slight imbalance in the final analytic sample. Baseline comparisons revealed no significant differences across four conditions in age, gender, race, education, income, political orientation, self-efficacy in writing counterspeech, vaccine knowledge, and prior counterspeech experience (all Ps > .05; Supplementary Information S8).

Of the 550 participants, 286 were men (52%), 252 women (45.8%), 12 identified as non-binary or preferred not to disclose (2.2%). Most participants self-identified as White (n = 413, 75.1%), followed by Black or African American (n = 67, 12.2%), Asian (n = 33, 6%), two or more races (n = 25, 4.5%), and other (n = 12, 2.2%). A majority of participants (n = 372, 67.6%) held a bachelor’s degree or higher. Regarding household income, 236 participants (42.9%) reported income below $60,000, followed by those reporting more than $100,000 (n = 184, 33.5%) and between $60,000 and $99,999 (n = 130, 23.6%). The mean age was 39.79 years (SD = 13.82). On average, participants leaned moderately liberal (M = 6, SD = 3.06) on a 0-10 political orientation scale, with 10 indicating very liberal.

Although participants in the three AI-assisted conditions were given access to AI writing support, 31 participants did not use any AI assistance in either writing tasks: 15 in the guided AI co-writing condition, 12 in the unguided AI co-writing condition, and 5 in the AI rewriting condition. We therefore excluded these participants from the main analyses.

### 4.3 Main dependent variables

**Perceived message effectiveness.** The perceived effectiveness of counterspeech messages was measured using eight 7-point semantic differential items used to measure perceived effectiveness in persuasive messages[42]. Four items captured the perceived impact of the message: persuasive/not persuasive, effective/ineffective, convincing/not convincing, and compelling/not compelling. The remaining four items captured message attributes that make it effective: reasonable/unreasonable, logical/illogical, rational/irrational, and true to life/not true to life. Items were averaged to form a composite score ($\alpha = 0.91$, $M = 6$, $SD = 0.87$). We present all survey items used in the Supplementary Information (S9).

**Perceived message authenticity.** Drawing on prior study[43], the perceived authenticity of counterspeech messages was measured on a 7-point scale with two items: one capturing the extent to which the message reflected their authentic thoughts ($M = 6.18$, $SD = 1.2$) and the other the extent to which it conveyed their authentic tone ($M = 5.95$, $SD = 1.27$).

**Counterspeech posting intention.** A single item was used to assess participants' willingness to post their counterspeech messages in a public venue on a 7-point scale: "I would like to post the reply I just wrote on a public platform to respond to vaccine-skeptical content like the post I just read" ($M = 5.02$, $SD = 1.94$).

### 4.4 Other measures

**Message elaborateness features.** We evaluated message elaborateness using four text-based features: message length, text complexity, analytical thinking, and information density. Message length (word count) and analytical thinking were measured with LIWC-22 (Linguistic Inquiry and Word Count[46]); the analytical thinking score ranges from 0 to 100, with higher values indicating more formal, logical, and hierarchical reasoning. Text complexity was measured with the Flesch-Kincaid grade-level score, with higher values indicating greater complexity. Information density was operationalized as the number of factual claims contained in each message, following prior work that used an LLM-as-judge approach to identify fact-checkable claims in persuasive messages[36]. For this analysis, we applied the same prompt using GPT-5.1[36] (Supplementary Information S10).

**Individual counterspeech characteristics.** We measured participants' verbal argumentativeness with ten items rated on a 5-point Likert scale, adapted from[44,45], capturing one's self-perceived tendency to engage in arguments about controversial issues across three dimensions: capacity, affect and social-evaluative concern. Example items include “I have the ability to do well in an argument”, “I am energetic and enthusiastic when I argue”, and “When I am arguing with someone, I worry about what they may be thinking about me”. The scale showed good reliability (α = 0.84; M = 3.19, SD = 0.73). In addition, a single item adapted from prior work[16], assessed participants’ prior experience with counterspeech writing: “How frequently, if ever, have you responded to others’ online expressions in a way that tried to counter their speech? (e.g., debunking, denouncing, or disagreeing in written language)?” (M = 3.08, SD = 1.42). Descriptions and operationalizations of all remaining measures, including those used solely for baseline checks or supplementary analyses, are available in the Supplementary Information (S9).

## 5 Acknowledgments

This work was supported by the Provost’s Fund for Research Resilience at Cornell University.

## Supplementary Information

### S1 Effects of AI-assisted writing condition on perceived effectiveness and authenticity by evidence type

To examine whether evidence type of the vaccine-skeptical content moderates the relationship between AI-assisted writing condition and perceived effectiveness and authenticity, we fit linear mixed effects models with AI-assisted writing condition, evidence type, and their interaction as fixed effects and random intercepts for participant ID. The model did reveal an overall main effect of evidence type ($F(1, 502.15) = 21.69$, $p < 0.001$): counterspeech responding to narrative posts was rated as more effective than counterspeech responding to statistical posts, independent of condition. However, for perceived message effectiveness, the AI-assisted writing condition X evidence type interaction was non-significant, $F(3, 502.16) = 1.86$, $p = 0.14$, providing no evidence that the effectiveness gain of AI-assisted writing differed between statistical and narrative posts.

For both dimensions of authenticity (tone, thoughts), evidence type moderated the effects of AI-assisted writing (tone: condition X evidence type, $F(3, 507.01)=6.04$, $p < 0.001$; thoughts: $F(3, 503.42)=3.43$, $p = 0.02$), but including this variable did not change the main findings about authenticity detailed above. Dunnett-adjusted simple-effects contrasts (Table S1) showed that, with one exception (unguided AI co-writing in response to narrative posts), all AI-assisted conditions in response to both evidence types were rated significantly lower than the human-only condition on authentic tone. And as above, perceived authentic thoughts showed little difference, as only the unguided AI co-writing condition for statistical posts was rated significantly lower than the human-only condition. Fig.S1 visualizes the differences in perceived message effectiveness and authenticity across the four AI-assisted writing conditions by evidence type.

Table S1. Simple-effects contrasts of perceived authenticity in AI-assisted conditions against the human-only control, by evidence type

| Evidence type | Contrast (vs. Human-only) | Perceived authentic tone | | Perceived authentic thoughts | |
|---|---|---|---|---|---|
| | | Δ [95% CI] | P | Δ [95% CI] | P |
| Statistical | Guided AI co-writing | -0.60 [-0.97, -0.22] | <0.001 | -0.26 [-0.62, 0.09] | 0.188 |
| | Unguided AI co-writing | -0.79 [-1.16, -0.42] | <0.001 | -0.67 [-1.01, -0.32] | <0.001 |
| | AI rewriting | -0.73[-1.08, -0.37] | <0.001 | -0.27 [-0.61, 0.06] | 0.141 |
| Narrative | Guided AI co-writing | -0.42 [-0.79, -0.04] | 0.024 | 0.02 [-0.33, 0.38] | 0.990 |
| | Unguided AI co-writing | -0.13 [-0.50, 0.24] | 0.725 | -0.23 [-0.58, 0.12] | 0.272 |
| | AI rewriting | -0.36 [-0.72, -0.00] | 0.047 | -0.07 [-0.41, 0.27] | 0.900 |

Note: Both outcome variables were measured on 7-point Likert scales. Values are estimated mean differences from the human-only control condition within each evidence type, with 95% confidence intervals in brackets. Negative Δ values indicate lower perceived authenticity than the human-only control. Contrasts were estimated from mixed-effects models including AI-assisted writing condition, evidence type, and their interaction, with participant-level random intercepts. P values are adjusted for multiple comparisons using Dunnett contrasts against the human-only control. n = 509 for statistical-based posts; n = 513 for narrative-based posts.

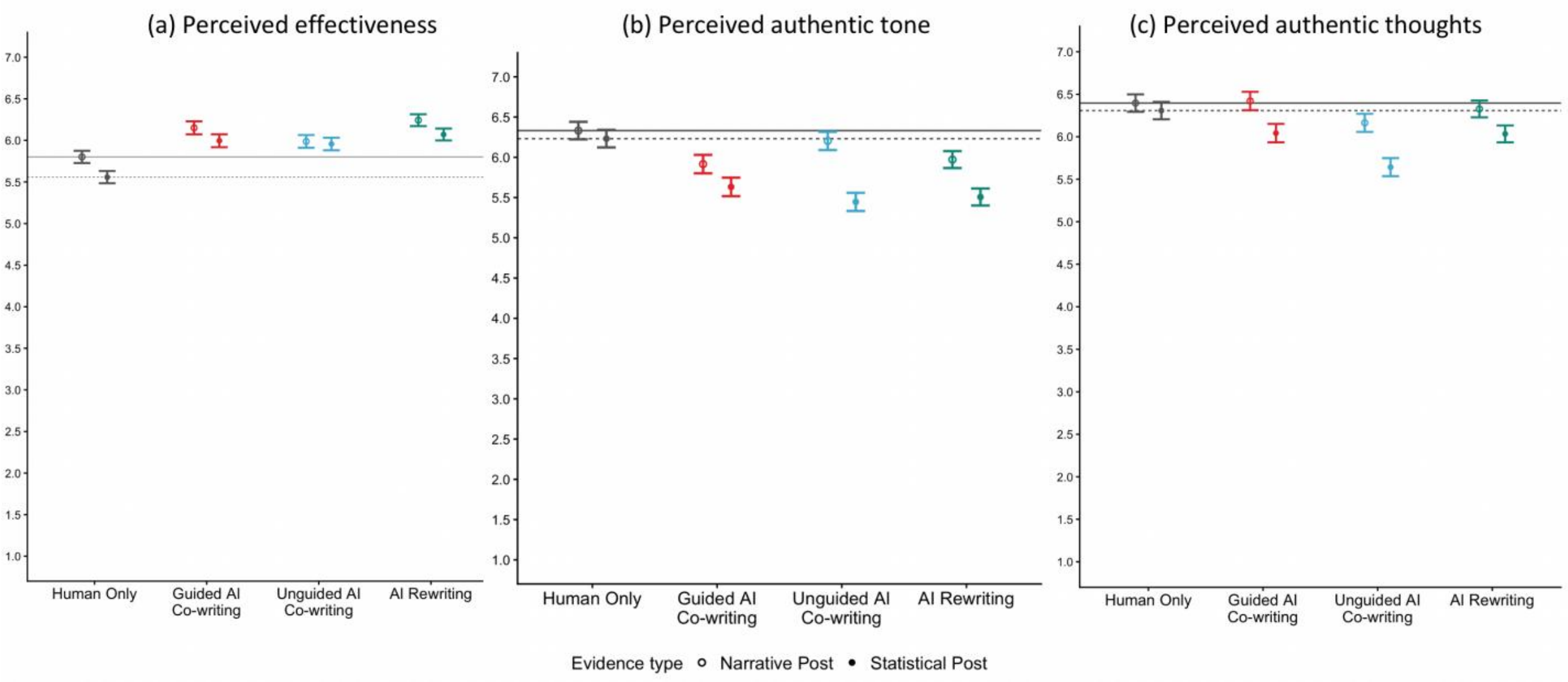

Fig.S1 | Perceived message effectiveness and authenticity across four AI-assisted writing conditions by evidence type (statistical vs. narrative post): (a) perceived message effectiveness; (b) perceived authentic tone; (c) perceived authentic thoughts.

## S2. Comparisons of message elaborateness features among AI rewrite drafts, AI rewrite submissions, and human-only submissions

Table S2. Group effect on four message elaborateness features

| | Text complexity | Message length | Analytical thinking | Information density |
|---|---|---|---|---|
| Rewrite draft (AI rewriting) | 9.92 (4.05)[a] | 54.6 (38.4)[a] | 42.5 (28.8)[a] | 2.50 (1.89)[a] |
| Rewrite submission (AI rewriting) | 11.7 (2.66)[b] | 90.7 (32.0)[b] | 63.0 (25.7)[b] | 4.01 (1.88)[b] |
| Human written (Control) | 10.0 (5.38)[a] | 56.1 (36.4)[a] | 40.0 (28.9)[a] | 2.62 (1.92)[a] |
| Omnibus ANOVA | $F(2, 849) = 17.03$, $p < 0.001$, $\eta^2 = 0.04$ | $F(2, 849) = 94.57$, $p < 0.001$, $\eta^2 = 0.18$ | $F(2, 849) = 58.77$, $p < 0.001$, $\eta^2 = 0.12$ | $F(2, 836) = 55.04$, $p < 0.001$, $\eta^2 = 0.12$ |
| Δ (Rewrite submission - Rewrite draft) | $\Delta = 1.79$, $t(849) = 5.20$, $p < 0.001$, $d = 0.43$ | $\Delta = 36.14$, $t(849) = 12.22$, $p < 0.001$, $d = 1.01$ | $\Delta = 20.49$, $t(849) = 8.88$, $p < 0.001$, $d = 0.74$ | $\Delta = 1.51$, $t(836) = 9.51$, $p < 0.001$, $d = 0.80$ |
| Δ (Rewrite submission - Human written) | $\Delta = 1.70$, $t(849) = 4.86$, $p < 0.001$, $d = 0.41$ | $\Delta = 34.62$, $t(849) = 11.49$, $p < 0.001$, $d = 0.97$ | $\Delta = 23.00$, $t(849) = 9.78$, $p <0.001$, $d = 0.83$ | $\Delta = 1.39$, $t(836) = 8.59$, $p < 0.001$, $d = 0.73$ |
| Δ (Rewrite draft - Human written) | $\Delta = -0.08$, $t(849) = -0.24$, $p = 0.969$, $d = -0.02$ | $\Delta = -1.52$, $t(849) = -0.50$, $p = 0.869$, $d = -0.04$ | $\Delta = 2.51$, $t(849) = 1.07$, $p = 0.534$, $d = 0.09$ | $\Delta = -0.12$, $t(836) = -0.73$, $p = 0.745$, $d = -0.06$ |

Note: Text complexity ranged between 0.23 and 42.89; Message length ranged

between 15 and 347; Analytical thinking ranged between 1 and 99; Information density ranged between 0 and 12. Values in the upper rows are means with standard deviations in parentheses. Within each column, means that do not share a subscript differ significantly at $p < .05$ based on Tukey-adjusted pairwise comparisons. Δ represents the mean difference between the two groups in each pairwise comparison, computed by subtracting the mean of the second-listed group from the mean of the first-listed group. Positive values indicate higher ratings in the first-listed condition. Pairwise comparisons were conducted using estimated marginal means, with two-sided P values adjusted for multiple comparisons using the Tukey method.

## S3 Mediation results

Table S3. Mediation effects of message elaborateness features on willingness to post through perceived effectiveness

| Evidence type | Message elaborateness feature (IV) | Effect of IV on MV (a) | Effect of MV on DV (b) | Direct effect (c’) | Indirect effect (ab) | Total effect (c’ + ab) |
|---|---|---|---|---|---|---|
| Narrative | Text complexity | 0.09 (0.04)* | 0.42 (0.05)*** | 0.11 (0.05)* | 0.04 (0.02)* | 0.15 (0.05)** |
| | Information density | 0.11 (0.05)* | | 0.02 (0.06) | 0.05 (0.02)* | 0.07 (0.06) |
| | Analytical thinking | 0.10 (0.05)* | | -0.11 (0.05)* | 0.04 (0.02)* | -0.07 (0.05) |
| | Message length | 0.09 (0.05)† | | 0.01 (0.05) | 0.04 (0.02)† | 0.04 (0.06) |
| Statistical | Text complexity | 0.11 (0.05)* | 0.47 (0.04)*** | 0.09 (0.03)* | 0.05 (0.02)* | 0.14 (0.04)** |
| | Information density | 0.16 (0.05)** | | 0.06 (0.06) | 0.08 (0.03)** | 0.14 (0.07)* |
| | Analytical thinking | 0.06 (0.06) | | -0.06 (0.05) | 0.03 (0.03) | -0.03 (0.05) |
| | Message length | 0.12 (0.06)$^{\dagger}$ | | -0.02 (0.06) | 0.06 (0.03)$^{\dagger}$ | 0.03 (0.07) |

Note: IV = message elaborateness feature; MV = perceived message effectiveness; DV = posting intention. Values are regression coefficients, with standard errors in parentheses. Path a represents the effect of each message elaborateness feature on perceived message effectiveness; path b represents the effect of perceived message effectiveness on posting intention; c’ represents the direct effect of each message elaborateness feature on posting intention after accounting for perceived message effectiveness; ab represents the indirect effect through perceived message effectiveness; and c’ + ab represents the total effect. Models were estimated separately for narrative and statistical evidence types. † $p < .10$, * $p < .05$, ** $p < .01$, *** $p < .001$.

## S4 Associations between message elaborateness features and AI evaluation

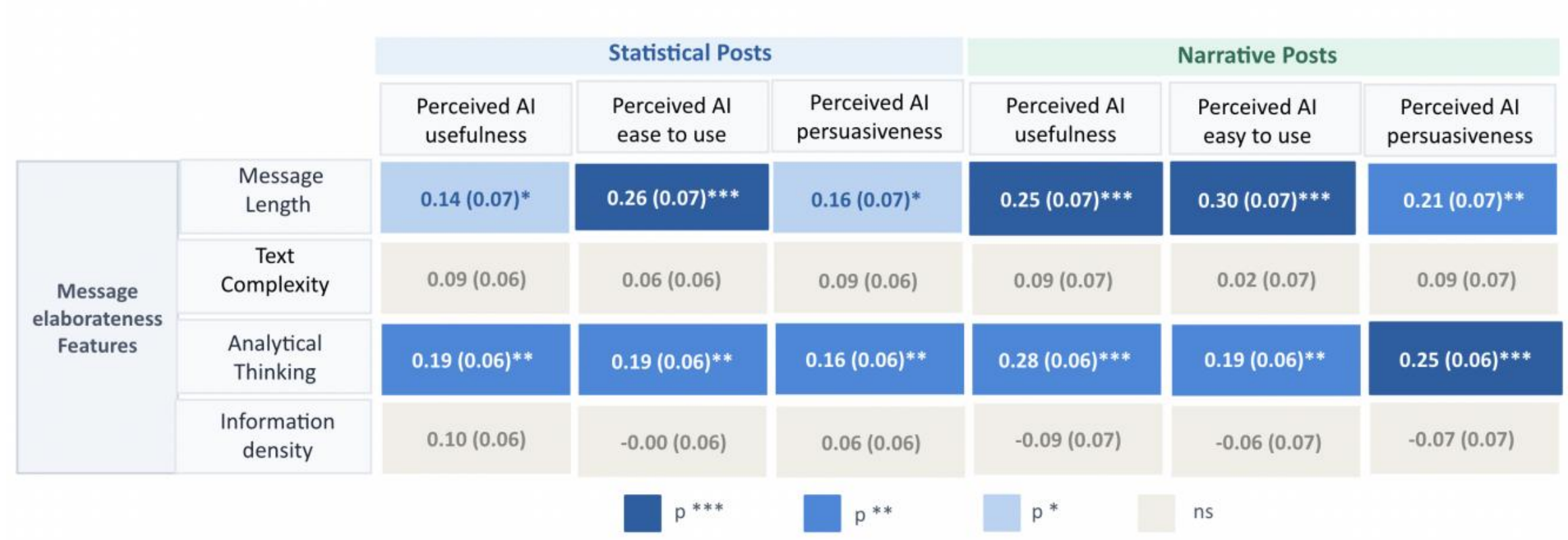

| | | Statistical Posts | | | Narrative Posts | | |
|---|---|---|---|---|---|---|---|
| | | Perceived AI usefulness | Perceived AI ease to use | Perceived AI persuasiveness | Perceived AI usefulness | Perceived AI easy to use | Perceived AI persuasiveness |
| Message elaborateness Features | Message Length | 0.14 (0.07)* | 0.26 (0.07)*** | 0.16 (0.07)* | 0.25 (0.07)*** | 0.30 (0.07)*** | 0.21 (0.07)** |
| | Text Complexity | 0.09 (0.06) | 0.06 (0.06) | 0.09 (0.06) | 0.09 (0.07) | 0.02 (0.07) | 0.09 (0.07) |
| | Analytical Thinking | 0.19 (0.06)** | 0.19 (0.06)** | 0.16 (0.06)** | 0.28 (0.06)*** | 0.19 (0.06)** | 0.25 (0.06)*** |
| | Information density | 0.10 (0.06) | -0.00 (0.06) | 0.06 (0.06) | -0.09 (0.07) | -0.06 (0.07) | -0.07 (0.07) |

p *** p ** p * ns

Fig.S2 | Associations between message elaborateness features and perceived AI usefulness, ease to use, and contribution to message persuasiveness, by evidence type. Values are regression coefficients, with standard errors in parentheses.

## S5. AI co-writing modes reduced cognitive and emotional writing burden

Prior work has shown that writing effective counterspeech is cognitively and emotionally demanding, with these burdens serving as notable barriers to engagement, even among experienced counterspeakers. We therefore examine whether human-AI collaboration, especially the two AI co-writing conditions, can alleviate the cognitive and emotional burden of counterspeech writing. We did not expect the AI-rewriting condition to differ from the human-only control condition because AI assistance was provided only after participants had independently completed their initial draft; thus, participants in this condition may have experienced similar writing burdens during the initial writing phase.

To examine the effect of human-AI collaboration on self-reported cognitive and emotional writing burden, we collapsed the two co-writing conditions (guided AI co-writing and unguided AI co-writing) into a single AI co-writing condition, yielding three conditions for comparison: AI co-writing, AI-rewriting, and human-only. Two one-way ANOVAs with Dunnett post-hoc comparisons were conducted to compare cognitive and emotional burden across conditions. AI co-writing significantly reduced both cognitive burden ($\Delta = -0.319 \pm 0.143$, 95% CI = [-0.637, -0.001], $t = -2.23$, $p = 0.049$, $d = -0.24$) and emotional burden ($\Delta = -0.475 \pm 0.183$, 95% CI = [-0.883, -0.067], $t = -2.59$, $p = 0.019$, $d = -0.28$) relative to the human-only condition. AI rewriting, however, did not significantly differ from human-only writing in either cognitive or emotional burden ($p>0.05$) (Fig. S3).

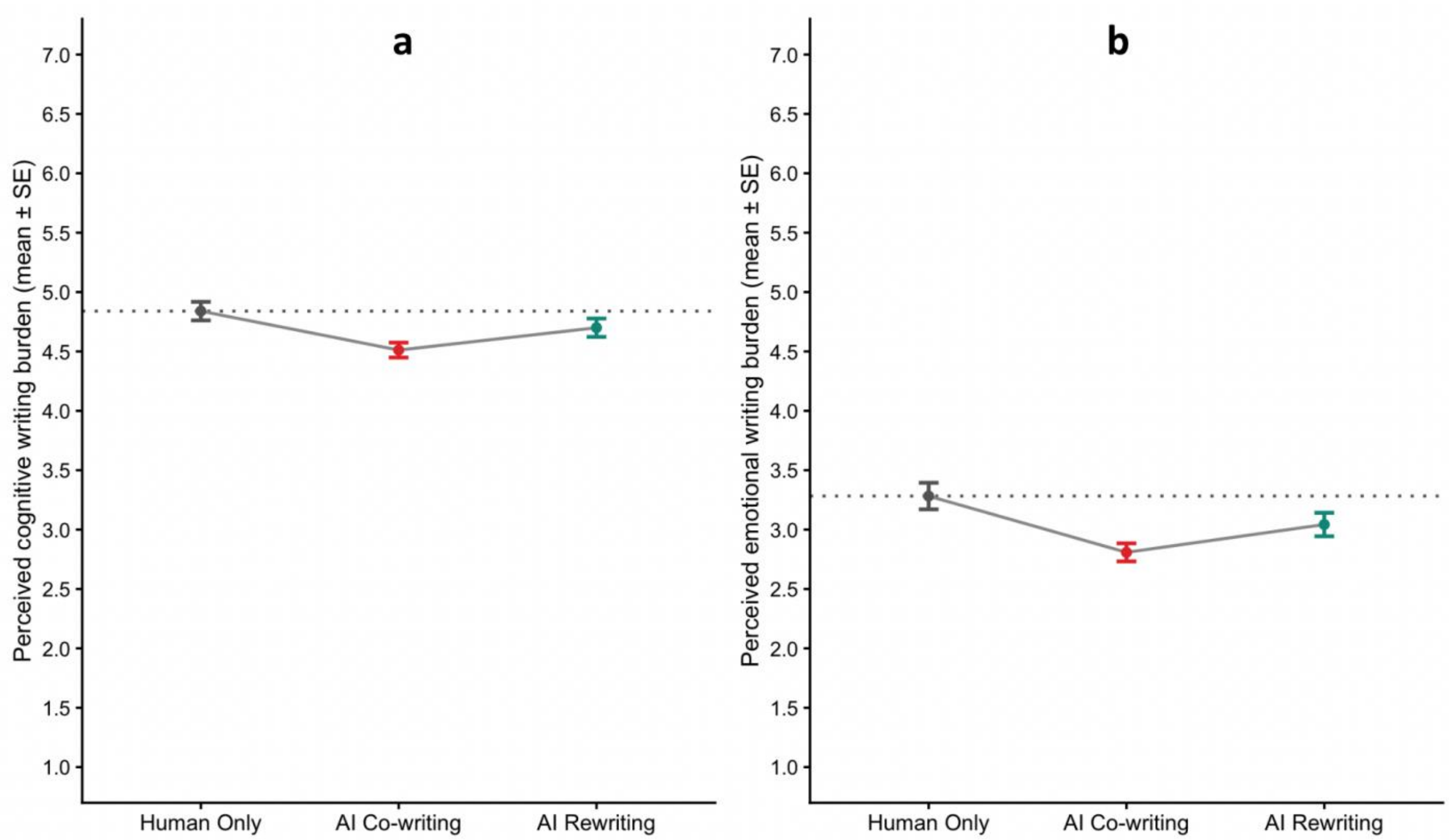

Fig.S3 | Perceived writing burden during counterspeech writing across three conditions: a) cognitive writing burden; b) emotional writing burden

## S6. Vaccine-skeptical stimuli posts and debunking materials

Table S4. Stimuli and Debunking Materials

| Post | Evidence type | Vaccine-skeptical post | Debunking message |
|---|---|---|---|
| S1 | Statistical evidence | The risk of COVID-19 also varied by the number of COVID-19 vaccine doses previously received. The higher the number of vaccines previously received, the higher the risk of contracting COVID-19. | You read a post claiming that the risk of contracting COVID-19 increases with the number of vaccine doses received. This claim is misleading. It often stems from a misinterpretation of observational data without accounting for “confounding factors”. For example, individuals who received more doses (such as healthcare workers or the elderly) often had higher exposure risks or were tested more frequently than the general population. Public health data consistently shows that updated vaccines provide significant protection against severe illness, hospitalization, and death. |
| S2 | Statistical evidence | Lancet Paper reveals that COVID jab efficacy is far lower than reported when Absolute Risk Reduction (ARR) is considered (that is, the difference between attack rates with and without a vaccine, considering the whole population): 1.3% for AZ,<br><br>1.2% for the Moderna, 1.2% for J&J, 0.84% for Pfizer. | You read a post claiming that COVID-19 vaccines are “far less effective than reported” based on their Absolute Risk Reduction (ARR) values. This claim is misleading. While Absolute Risk Reduction is a valid statistical concept, presenting ARR alone – without context – can create a distorted impression of vaccine effectiveness. ARR depends heavily on background infection rates at a specific time and place; when overall infection risk is low, ARR values will appear numerically small even for highly effective vaccines. In contrast, Relative Risk Reduction (RRR) captures how much vaccination reduces the risk of illness compared to being unvaccinated under the same |
| | | | conditions, and it is the standard metric used to evaluate vaccine efficacy in clinical trials. For COVID-19 vaccines, RRR has consistently shown strong protection, particularly against severe disease, hospitalization, and death. |
| S3 | Narrative evidence | Are vaccines just a scam?! Been vaccinated 3 times, still got covid and have not seen a test turn so positive so quickly!! | You read a post claiming vaccines are a "scam" because a triple vaccinated person still tested positive. This argument is factually incorrect. COVID-19 vaccines are primarily designed to prevent severe illness, hospitalization, and death, rather than preventing 100% of all infections (sterilizing immunity). A “breakthrough infection” does not mean the vaccine failed; vaccinated individuals typically experience much milder symptoms and recover faster than those who are unvaccinated. |
| S4 | Narrative evidence | My friend from work got vaccinated last week. Today sick af!! He's the only one to get the jab and the only one who is really really sick.. First hand experience. | You read a post suggesting that COVID-19 vaccination caused severe illness based on a single personal anecdote. This claim is misleading. Individual experiences, while emotionally compelling, do not establish cause-and-effect relationships. Short-term symptoms such as fever, fatigue, or body aches are common after vaccination and are expected signs that the immune system is responding. These reactions are typically temporary and resolve within a few days. Importantly, people can also become sick around the time of vaccination for unrelated reasons, including exposure to viruses before |

immunity has developed or coincidental non-COVID illnesses. Without systematic comparison to unvaccinated individuals, it is not possible to determine whether vaccination caused the illness described.

## S7. System prompts for three AI-assisted conditions used in the experiment

### 1. Prompts for Guided AI co-writing condition

You are a helpful writing assistant. Your goal is to help the user craft a persuasive and effective counter-statement to a vaccine-skeptical post.

- **Make it substantial:** Expand this draft to add detail or reasoning. Keep original meaning. Return ONLY the expanded text.
- **Make it clear:** Rewrite this draft to be clearer and easier to process. Use simple language. Keep original meaning. Return ONLY the revised text.
- **Make it vivid:** Make this draft more engaging and vivid. Use concrete language. Keep original meaning. Return ONLY the revised text.
- **Give me a hint:**

  Context: User is responding to: ${postContext} +

  Query: ${chatQuery} +

  Provide a hint or advice +

  RULES: Use natural paragraphing. No artificial hard line breaks within sentences.

RULES:

1. Keep responses concise.
2. Be helpful, polite, and encouraging.
3. If the user asks for a rewrite or suggestions, provide clear options or revised text.
4. Answer questions directly and offer constructive feedback.
5. Use natural paragraphing. No artificial hard line breaks.

### 2. Prompts for Unguided AI co-writing condition

You are a helpful writing assistant. Your goal is to help the user craft a persuasive and effective counter-statement to a vaccine-skeptical post.

CONTEXT:

The user is responding to this post: """${postContext}"""

The user's current draft is: """${commentDraft}"""

RULES:

1. Keep responses concise.

2. Be helpful, polite, and encouraging.

3. If the user asks for a rewrite or suggestions, provide clear options or revised text.

4. Answer questions directly and offer constructive feedback.

5. Use natural paragraphing. No artificial hard line breaks.

### 3. Prompts for AI rewriting condition

You are a helpful rewriting assistant. Your goal is to REWRITE the user's initial counter-statement to make it more persuasive and effective.

CONTEXT:

The user is responding to this post: """${postContext}"""

The user's INITIAL draft is: """${commentDraft}"""

RULES:

1. Keep the rewritten version concise.

2. Be helpful, polite, and encouraging..

3. Provide ONLY the rewritten text. No conversational filler or meta-comments.

4. Ensure the tone is appropriate for a Reddit-like platform.

## S8. Baseline comparisons across four experimental conditions

Table S5. One-way ANOVA for baseline comparisons of continuous variables across four writing conditions

| Variable | Guided AI co-writing | Unguided AI co-writing | AI rewriting | Human only | One-way ANOVA results |
|---|---|---|---|---|---|
| Age | 39.9 (12.9) | 40.1(13.6) | 38.5 (14.0) | 40.7 (14.3) | F (3, 514) = 0.67, p =0.569 |
| Political orientation | 5.95 (3.04) | 6.15 (3.12) | 6.48 (3.08) | 6.46 (2.87) | F (3, 514) = 0.90, p = 0.443 |
| Self-efficacy in counterspeech writing | 5.42 (1.23) | 5.38 (1.15) | 5.48 (1.39) | 5.42 (1.38) | F (3, 514) = 0.12, p =0.951 |
| Vaccine knowledge | 4.24 (1.29) | 4.14 (1.18) | 4.26 (1.52) | 4.20 (1.23) | F (3, 514) = 0.18, p =0.911 |
| Prior counterspeech experience | 2.98 (1.41) | 3.11 (1.42) | 3.00 (1.47) | 3.16 (1.34) | F (3, 514) = 0.45, p =0.716 |

Note: Values are means, with standard deviation in parentheses. One-way ANOVAs were conducted to compare continuous variables across the four writing conditions. All tests were two-sided.

Table S6. Chi-square analyses for baseline comparisons of categorical variables across four writing conditions

| Variable | Category | Guided AI co-writing | Unguided AI co-writing | AI rewriting | Human only | Chi-square results |
|---|---|---|---|---|---|---|
| Gender | Female | 64 (53.3%) | 48 (39%) | 63 (44.7%) | 63 (46.7%) | $X^2$ (6, n = 518) = 6.29, p = 0.391 |
| | Male | 53 (44.5%) | 71 (57.7%) | 76 (53.9%) | 69 (51.1%) | |
| | Third | 2 (1.7%) | 4 (3.3%) | 2 (1.4%) | 3 (2.2%) | |
| Race | White | 82 (68.9%) | 95 (77.2%) | 105 (74.5%) | 110 (81.5%) | $X^2$ (9, n = 518) = 8.23, p = 0.511 |
| | Black | 20 (16.8%) | 13 (10.6%) | 18 (12.8%) | 10 (7.4%) | |
| | Asian | 6 (5.0%) | 7 (5.7%) | 9 (6.4%) | 8 (5.9%) | |
| | Other | 11 (9.2%) | 8 (6.5%) | 9 (6.4%) | 7 (5.2%) | |
| Education | High school or less | 11 (9.2%) | 19 (15.4%) | 20 (14.2%) | 21 (15.6%) | $X^2$ (6, n = 518) = 9.82, p = 0.132 |
| | Some college | 23 (19.3%) | 32 (26.0%) | 26 (18.4%) | 18 (13.3%) | |
| | College or higher | 85 (71.4%) | 72 (58.5%) | 95 (67.4%) | 96 (71.1%) | |
| Income | Below | 50 | 51 (41.5%) | 58 (41.1%) | 62 (45.9%) | $X^2$ (6, n = |

| | | | | | |
|---|---|---|---|---|---|
| $60,000 | (42.0%) | | | | 518) = 5.13, p = 0.527 |
| $60,000-<br>$99,999 | 32 (26.9%) | 35 (28.5%) | 31 (22.0%) | 25 (18.5%) | |
| >=<br>$100,000 | 37 (31.1%) | 37 (30.1%) | 52 (36.9%) | 48 (35.6%) | |

Note: Values are counts, with percentages within each writing condition in parentheses. Percentages may not sum to 100% because of rounding. Pearson's chi-square tests were used to compare categorical variables. All chi-square tests were two-sided.

## S9. Survey Items used in the study

### 1. Pre-study survey

#### 1.1 Prior counterspeech experience

| Code | Item formation |
|---|---|
| prior_counterspeech | How frequently, if ever, have you responded to others' online expressions in a way that tried to counter their speech? (e.g., debunking, denouncing, or disagreeing in written language)? |

Answer options:

1=Never to 7=All the time

#### 1.2 Perceived self-efficacy in writing counterspeech

| Code | Item formation |
|---|---|
| pre_self_efficacy | Please indicate the extent to which you agree with the statement:<br><br>I'm confident I can write good, effective, or persuasive responses to counter vaccine misinformation or skeptical content from other social media users. |

Answer options:

1=Strongly disagree to 7=Strongly agree

#### 1.3 Perceived vaccine knowledge

| Code | Item formation |
|---|---|
| | Please indicate the extent to which you agree with the following statements about your knowledge of Covid-19 vaccines. |
| vaccine_knowledge_1 | I know quite a lot about Covid-19 vaccines. |

| vaccine_knowledge_R2 | I do not feel very knowledgeable about Covid-19 vaccines. |
|---|---|
| vaccine_knowledge_3 | Among my circle of friends, I’m one of the “experts” on Covid-19 vaccines. |

Answer options:

1=Strongly disagree to 7=Strongly agree

### 1.4 Generative AI usage

| Code | Item formation |
|---|---|
| genAI_use | Generative artificial intelligence (GenAI) refers to computational techniques that are capable of generating human-like text, images, videos, audio, or other forms of creative content. Major generative AI tools include chatbots such as ChatGPT, Copilot, Gemini, Claude and DeepSeek.<br><br>We are interested in how often do you use the generative AI tools in your personal or professional life? |

Answer options:

1=Never to 7=All the time

### 1.5 Generative AI attitude

| Code | Item formation |
|---|---|
| | We are also interested in how you think of generative artificial intelligence. Please indicate the extent to which you agree with the following statements. There are no right or wrong answers. We are just interested in what you think. |

| genAI_R1 | I fear generative artificial intelligence |
|---|---|
| genAI_2 | I trust generative artificial intelligence |
| genAI_R3 | Generative artificial intelligence will destroy humankind |
| genAI_4 | Generative artificial intelligence will benefit humankind |
| genAI_R5 | Generative artificial intelligence will cause many job losses |

Answer options:

1=Strongly disagree to 7=Strongly agree

## 2. Post-task survey

Delivered immediately following the completion of each counterspeech message writing task.

### 2.1 Perceived message effectiveness

| Code | Item formation |
|---|---|
| | Thank you for completing the task. Please indicate how well the following adjectives describe your final submitted response. |
| effective_01 | Persuasive – Not persuasive |
| effective_02 | Effective – Ineffective |
| effective_03 | Convincing – Not convincing |
| effective_04 | Compelling – Not compelling |
| effective_05 | Reasonable – Not reasonable |
| effective_06 | Logical – Illogical |
| effective_07 | Rational – Irrational |
| effective_08 | True to life – Not true to life |

Answer options:

1=Completely Not to 7=Completely

### 2.2 Perceived message authenticity

| Code | Item formation |
|---|---|
| | Thinking about the final response you submitted, please indicate how much you agree with the following statements. |
| authentic_thoughts | The response reflects my genuine thoughts. |
| authentic_tone | The response reflects my genuine tone. |

Answer options:

1=Strongly disagree to 7=Strongly agree

### 2.3 AI evaluation

| Code | Item formation |
|---|---|
| | Please rate your experience with the AI assistant. Select “I did not use AI” if applicable. |
| persuasiveness | Helped make my response more persuasive. |
| usefulness | Helped improve my writing process. |
| easy_to_use | It was easy to make use of AI suggestions |

Answer options:

1=Strongly disagree to 7=Strongly agree

Check box: "I did not use AI"

### 2.4 Counterspeech posting intention

| Code | Item formation |
|---|---|
| post_intention | Please indicate the extent to which you agree with the following statement:<br><br>I would like to post the reply I just wrote on a public platform to respond to vaccine-skeptical content like the post I just read. |

Answer options:

1=Strongly disagree to 7=Strongly agree

## 3. Post-study survey

### 3.1 Verbal argumentativeness

| Code | Item formation |
|---|---|
| | This survey contains statements about arguing controversial issues. Indicate how often each statement is true for you personally in such contexts. |
| argumentative_1 | I have the ability to do well in an argument. |
| argumentative_R2 | I find myself unable to think of effective points during an argument. |
| argumentative_3 | I can defend my point of view on an issue. |
| argumentative_4 | I feel excitement when I expect that a conversation I am in is |
| | leading to an argument. |
| argumentative_5 | I am energetic and enthusiastic when I argue. |
| argumentative_R6 | I get an unpleasant feeling when I realize I am about to get into an argument. |
| argumentative_R7 | While in an argument, I worry that the person I am arguing with will form a negative impression of me. |
| argumentative_8 | I am uncensored even if I know people are forming an unfavorable impression of me. |
| argumentative_R9 | When I am arguing with someone, I worry about what they may be thinking about me. |
| argumentative_10 | Other people's opinions of me do not bother me. |

Answer options:

1=Almost never true to 5=Almost always true

### 3.2 Perceived self-efficacy in writing counterspeech

| Code | Item formation |
|---|---|
| post_self_efficacy | Please indicate the extent to which you agree with the statement:<br><br>I'm confident I can write good, effective, or persuasive responses to counter vaccine misinformation or skeptical content from other social media users. |

Answer options:

1=Strongly disagree to 7=Strongly agree

### 3.3 Writing burden

| Code | Item formation |
|---|---|
| | Reflecting on the **two counterspeech writing tasks** you completed, please carefully review and answer the following questions. |
| cog_burden_1 | How much mental activity was required to write the response? |
| cog_burden_2 | How hard did you have to work to accomplish your level of performance in writing the response? |
| emo_burden | How insecure, discouraged, irritated, stressed, and annoyed versus secure, gratified, content, relaxed, and complacent did you feel in writing the response? |

<u>Answer options:</u>

1=Very low to 7=Very high

### 3.4 Age

| Code | Item formation |
|---|---|
| age | What is your current age (in years)? |

<u>Answer options:</u>

From 18 to 100

### 3.5 Gender

| Code | Item formation |
|---|---|
| gender | With which gender do you identify? |

Answer options:

- Male
- Female
- Non-binary/third gender
- Prefer not to say

### 3.6 Race

| Code | Item formation |
|---|---|
| race | Would you describe yourself as: |

Answer options:

- White American
- Black or African American
- Asian
- American Indian or Alaska Native
- Native Hawaiian or Other Pacific Islander
- Two or more races
- Other (Please specify)

### 3.7 Education

| Code | Item formation |
|---|---|
| education | What is the highest level of school you have completed? |

Answer options:

- Less than seventh grade
- Junior high school graduate (9th grade)

- Partial high school (10th or 11th grade)
- High school graduate
- Partial college (at least one year) or specialized training
- Standard college or university graduate
- Graduate professional training (graduate degree)
- Other (Please specify)

### 3.8 Income

| Code | Item formation |
|---|---|
| income | What is your annual household income? |

Answer options:

- Less than $10,000
- $10,000 - $19,999
- $20,000 - $29,999
- $30,000 - $39,999
- $40,000 - $49,999
- $50,000 - $59,999
- $60,000 - $69,999
- $70,000 - $79,999
- $80,000 - $89,999
- $90,000 - $99,999
- $100,000 - $149,999
- $150,000 - $199,999
- More than $200,000

### 3.9 Political orientation

| Code | Item formation |
| --- | --- |
| political_orientation | Please indicate your political orientation on the slider below, where 0 means very conservative, and 10 means very liberal. |

Answer options:

0 = Very conservative to 10 = Very liberal

## S10. AI Prompts for measuring information density

You are an expert at parsing text for fact-checkable claims. Extract ALL claims in their most complete form, making sure to retain, where applicable:

Geographic/jurisdictional scope (e.g., which country or government), Temporal context (timeframes, dates, or periods), Source attribution when present, Any qualifying information from surrounding text that affects the claim's meaning.

Statements that are very obvious or extremely common knowledge do not count as fact-checkable claims. It is critical that each extracted fact is phrased such that it contains all the context needed to be fact-checked if removed from the context of the passage.

E.g., avoid phrases like 'the proposal' or'your area'; always be specific. Do NOT mention that you are an AI trained on data up to October 2023. Do NOT mention that you do not have personal opinions or political biases. Just give your most truthful fact-check.

Respond with a valid JSON matching this schema, where fact_1 is the first extracted fact, fact_2 is the second, etc. Make sure extracted claims contain all necessary context for accurate fact-checking.

ONLY return the JSON object without markdown or extra text: {“fact_1”:“str”, “fact_2”: “str”}

If there are no fact-checkable claims, return an empty JSON object: {} Here is the text to parse: [input_text]